# Ringed spherulitic and undulated textures in the nematic phase of a mixture of trans-4-hexylcyclohexanecarboxylic and benzoic acids.

Amelia Carolina Sparavigna
Dipartimento di Fisica, Politecnico di Torino, C.so Duca degli Abruzzi 24, Torino, Italy

Abstract
We present a polarising optical microscopy study of a liquid-crystal mixture of trans-4-hexylcyclohexanecarboxylic acid (C6) and benzoic acid. Both materials have carboxylic groups that can form dimers through hydrogen bonding. The mixture is nematic and room temperature and has the clearing point at 88°C. The nematic phase shows ringed spherulites, sometimes looking like spiral structures, and undulated textures, which remain faintly visible heating the sample till 71 °C. This is the temperature of a texture transition inside the nematic phase.

1. Introduction
Compounds with aromatic carboxylic groups that can form dimers through hydrogen bonding are well-known examples of materials where the formation or stabilization of liquid crystalline phases is strictly due to dimerization. The study of these compounds is old [1-4], but new classes of compounds have recently been synthesized, the liquid crystalline behaviour of which is also dependent on intermolecular hydrogen bonds between similar or dissimilar molecules [5]. Moreover, interesting effects arise in well-known and studied materials, such as the appearance of textures transitions and of chiral domains [6-8]. Let us remember that hydrogen bonds are interactions between molecules where a hydrogen atom attached to an electronegative atom, through a polar covalent bond of one molecule, aligns with the unshared lone electron pair of an electronegative atom on another neighbouring molecule. Although the strongest hydrogen-bonding interaction is only about 12% of a covalent chemical bond, hydrogen bond is significantly stronger than dipole-dipole and dispersion forces. This interaction is then able to produce dimers, stable enough to create mesogenic molecules.
We have previously investigated some alkyloxybenzoic acids and cyclohexane-carboxylic acids and observed a texture transitions in the nematic phase and periodic instabilities at the phase transitions [9]. Alkyloxybenzoic and cyclohexane-carboxylic acids have smectic C and smectic B phases respectively. Here we discuss the mixture of trans-4-alkylcyclohexanecarbolxylic acid, hereafter called C6 (monomer structural formula $C_6H_{13}C_6H_{10}COOH$), with the benzoic acid. Fig.1 shows a C6 monomer (a) and two dimers (b),(c), a heterodimer of C6 monomer with a benzoic acid monomer of the benzoic acid and a homodimer with two C6 monomers. The existence of heterodimers is acceptable, considering the possibility to have monomers of benzoic acid in solution [10]. Assuming C6 as a solvent, we can guess that C6 homodimers and heterodimers of C6 and benzoic acid arise in the mixture. We cannot exclude that a small amount of benzoic acid could remains in its dimerized form. Note in Fig.1 the bent shape of heterodimers. We use the arrows in the figure as in Ref. [8], to have a better view of the three-dimensional structure of monomer and dimers.
We decided to study a C6 mixture because this material is able to develop spontaneous chiral domains in its nematic phase. This material is achiral but gains chiral properties due to the formation of oligomers with screw molecular structures. The binary mixture of a cyclohexane-carboxylic acid with benzoic acid under study is displaying interesting ringed spherulitic and undulated structures. The spherulitic structures look like spirals. The material possesses also a texture transition in its nematic phase.

## 2. The trans-4-hexyl-cyclohexane-carboxylic acid

Trans-4-hexyl-cyclohexane-carboxylic acids are cyclohexane rings with a carboxylic acid group and a hexyl tail affixed. The cyclohexane is a six-vertex ring, which does not conform to the shape of a perfect hexagon. The conformation of a flat two-dimensional planar hexagon has considerable angle strain and then to reduce this strain, trans-cyclohexane adopts a three-dimensional structure known as the chair conformation. This conformation puts the carbons at an angle of 109.5°. Therefore, in the cyclohexane, half of the hydrogen atoms are in the plane of the ring (equatorial position) while the other half are perpendicular to this plane (axial position). The conformation allows for the most stable structure of cyclohexane. Another conformation of cyclohexane exists, the cis-cyclohexane, known as boat conformation, but it converts to the more stable chair formation. If hydrogen is substituted with a large substituent, the substituent will most likely be found attached in an equatorial position, to have a more stable conformation. In C6, we have two substituent groups, an alkyl tail and a carboxylic group.

Let us shortly discuss two interesting properties displayed by C6 acid. The first is the presence of a spontaneous twist at certain conditions. In the experiment reported in Ref.8, spontaneously twisted domains appear on cooling from the isotropic phase, below but very close to the clearing point. In the cell, the researches observed few twisted domains, occurring randomly in different heating-cooling cycles. Their handedness was randomly distributed. The results of experiment tell that the formation of the chiral structure is definitely not due to anchoring, that is a surface effect, but to the generation of the nematic phase from intermolecular hydrogen bonds. In the isotropic phase, monomers are favoured. On cooling, closed and open dimers arise, creating the nematic phase. Below the clearing point, the interaction among open dimers and monomers can create oligomers, which have a screw shape and acting then as chiral dopants. The spontaneous twist has been then generated in a material where molecules are achiral: this fact is no more surprising now, because many materials with achiral molecules have been discovered, displaying spontaneous separation into left- and right-handed domains [11-16].

The second property of C6 is the presence of an order transition in its nematic phase. This transition, very smooth was observed on both cooling and heating the sample. The nematic sub-phase below the order transition, displays a texture with small and regular domains, different from that possessed by the nematic sub-phase above the transition, which is a Schlieren one. The presence of cybotactic clusters with local smectic order is usually viewed as the reason for a different order in the nematic melt. The nematic sub-phase at lower temperatures is then a cybotactic nematic. DeVries proposed a model for the arrangement of molecules in the cybotactic clusters, which looks like the arrangement of the smectic C phase [1].

In the usual classification of liquid crystals, the structure of a nematic is assumed lacking of lamellar order and featuring only long-range orientational order. A common picture is to imagine long rod-like molecules having a parallel mutual orientation, with the molecular centres of gravity arranged randomly without long-range correlation. However, many compounds have extremely similar x-ray diffraction patterns in both the nematic and smectic phases, implying that elements of local pseudo-lamellar ordering may exist in the nematic phase [17]. This hypothesis is in agreement with a model proposed in Ref.18, different from that of DeVries, to explain the x-ray scattering of cybotactic nematic. The model proposes a layered arrangement of molecules as that observed in the cholesteric nematic, without its twist. This lamellar model of cybotactic clusters gives a better agreement with x-ray observations; it is also good for C6, which does not have a smectic C phase, but a smectic B phase.

The existence of cybotactic clusters in liquid crystal compounds is receiving an increasing interest. We have recently observed a texture transition in the nematic phase of an oxadiazole compound [19,20]. We proposed the existence of cybotactic clusters to explain the spherulitic nematic phase in of this oxadiazole compounds. The existence of cybotactic clusters in another oxadiazole compound has been after reported [21]. The material has a ferroelectric response to a switching electric field in its nematic phase. The researchers concluded from simulations and x-ray observations that this ferroelectric response is due to a field-induced reorganization of polar cybotactic groups within the nematic phase. Cybotactic nematics are then interesting for progresses toward the realization of ferroelectric fluids.

Of course, from polarising optical microscopy (POM) observations we can suggest a possible cybotactic arrangement, in the case of materials with texture transitions. But, to infer any conclusion with respect to the molecular organisation in a mesophase, it would be better that optical studies were complemented by x-ray diffraction analysis, as in Ref.21. With both techniques, it is possible to identify and characterise the mesophase structure.

3. Sample preparation and POM observations

Here we use C6 and benzoic acids without further purification. C6 has the following phase sequence: Crystal-32°C-Smectic B-47°C-Nematic-96°C-Isotropic. Droplets of the mixture (90% in weight of C6 and 10% of benzoic acid) were deposited on untreated glasses. The glass surfaces were rubbed with cotton wool to favour a planar alignment. This procedure for inducing a slight planar alignment was suggested by researchers of the Liquid Crystal Group at Niopik Institute of Moscow, who referred it as useful in the case of thin cells. Over the droplets, a thin cover glass was placed. No spacer was used. We obtained three very thin samples and a sample rather thick. Let us estimate the thicknesses of thin films to be of few microns. The thick sample has with a rough estimate a thickness of 20 microns. Cells were heated and cooled in a Mettler–Toledo thermostage and textures observed with polarised light microscopy (POM). To determine the phase transition temperatures the scanning rate was set to 2 °C/min. The mixture is nematic at room temperature (28°C) and its clearing point is at 88°C.

The thick film displayed many domains as those shown in Fig.2: they could be ringed spherulitic structures, distorted by the presence of defects, and then looking like spirals. We have already observed spherulitic patterns in the nematic phase of oxadiazoles. In that case, on cooling the sample, it was possible to observe the transition of these nematic structures in the toric domains of a smectic phase. In the case of the mixture C6/benzoic acid, the transition to the smectic phase was not observed, because it is under our room temperature. We investigated then the samples increasing the temperature and saw that these spiral-like textures remain visible until 70°C. Over this temperature, no faint traces of the texture are visible. Fig.3 shows other spherulites and an interference pattern between such structures. The following Fig.4 reports details of the centre of these domains.

The structures in Fig.2 and 3 look like those observed in some polymeric materials. Spherulites are the basic morphology for polymers crystallization form melt, and they are usually observed in industrial processes. When observed by optical microscope between crossed polarizers, a cross-like extinction occurs. Ringed (or banded) spherulites have another extinction pattern, which consists in concentric bands superimposed. It is believed that the periodic extinction of ringed spherulites is led by a twisting along the radial direction during the crystal formation [22-25].

The patterns shown by Fig.2 and 3 have a very large size and are clearly visible. In other area of the cell these spherulitic structures are faintly visible. Heating the sample over the clearing point and cooling it at room temperature, these structures reappear in different places. It is necessary to observe that they are different from the spiral domains of cholesteric liquid crystals [26] and that the thin samples do not display them. In fact, if we squeeze the thick cell, the amount of liquid crystals reduces, and only few spherulitic domains remain at one side of the cell. We could conclude that it is possible a certain role of the thickness in the development of these textures.

The thin samples have Schlieren textures, which are quite interesting because they look like the cholesteric fingerprint textures. Figure 5 shows microphotography images of these patterns. The following figure (Fig.6) shows how the texture changes during the rotation of the sample. It is possible to have a better view for what concerns the undulation of the texture. This undulation is approximately of 10 microns and it is not a pre transitional structure appearing near the smectic phase, because it remains visible on heating the sample. Of course, it could be interesting to see what happens on cooling at the transition in the smectic phase, but we had not a set-up ready for this investigation.

A thin sample allows a better investigation of the existence of a texture transition in the nematic phase. The sample has usually a texture full of defectives, which looks like cholesteric texture. On heating, above 71°C, these defects disappear, leaving a smooth texture as shown in Fig.7. On cooling, defects reappear below 70°C.

4. Discussion

We know that C6 has an order transition in its nematic phase, probably due to the presence of cybotactic clusters. Again, we can consider the nematic below the texture transition as a cybotactic nematic. In the framework of model proposed in Ref.18, the clusters could have a lamellar order, as the cholesteric nematics have. Cholesteric nematics are mixtures of rod-like molecules with chiral dopants: here we have a binary nematic mixture too, obtained from C6 and benzoic acid. These materials have carboxylic groups that can form dimers through hydrogen bonding. As discussed in the introduction, we can imagine the nematic phase of this mixture as a melt of C6 homodimers with heterodimers of C6 and benzoic acid monomers, that is an assembly of rod-like homodimers with heterodimers acting as dopants, in a certain manner, due to their different length and shape. As shown in Fig.1, the heterodimers have a bent-shape. The textures that we observed in the nematic low temperature sub-phase could be due to specific properties of cybotactic clusters, the molecular organization of which is governed by the presence of bent-shape heterodimers.

We have then decided to investigate the role of concentration of heterodimers on the nematic texture. We have prepared another mixture of trans-4-hexylcyclohexanecarboxylic acid (C6) with a smaller percentage of benzoic acid (5% in weight). The mixture has the clearing point at 92°C. The texture slowly changes on cooling the sample but does not display a clear transition in a smectic phase or a texture transition. Below 42°C, the texture seems to have very small domains with zigzag arrangement. To have a better view of this texture, the sample is removed from the thermostage and observed by means of an objective with a very short focal length (this objective cannot be used to observe the cell in the thermostage). We see a striped texture with a very short period of 3 microns (see Fig.8), that could suggests that a lower concentration of benzoic acid means a shorter dimension of domains. The stripes are perpendicular to the rubbing direction and have a remarkable resembling with the textures of cholesteric nematics.

The microphotography shown by Fig.8 was obtained at room temperature (28°C): it was not possible to follow the texture on heating for the lack of resolution and then to conclude whether it is formed by pretransitional bars because the sample is reaching a smectic phase or it is a texture, as that shown in Fig.5, that remains stable at higher temperatures.

Here we have proposed some preliminary results on textures observed in binary mixture of C6 and benzoic acid. Future works are devoted to the investigation of the role of benzoic acid concentration on the behaviour of nematic domains.

Fig.1. Monomer (a) of C6, a heterodimer (b) with a benzoic acid and a homodimer with another C6 monomer. We use the arrows as in Ref. [8], to have a better view of the three-dimensional structure of monomer and dimers. Note that the heterodimer has a bent shape.

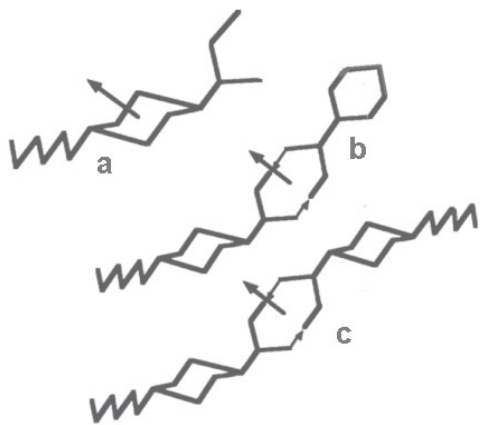

Fig.2. At room temperature the nematic phase of the mixture displays ringed spherulitic textures, which seems spirals. Note the large size of these structures. The mixture is 90% in weight of C6 and 10% of benzoic acid.

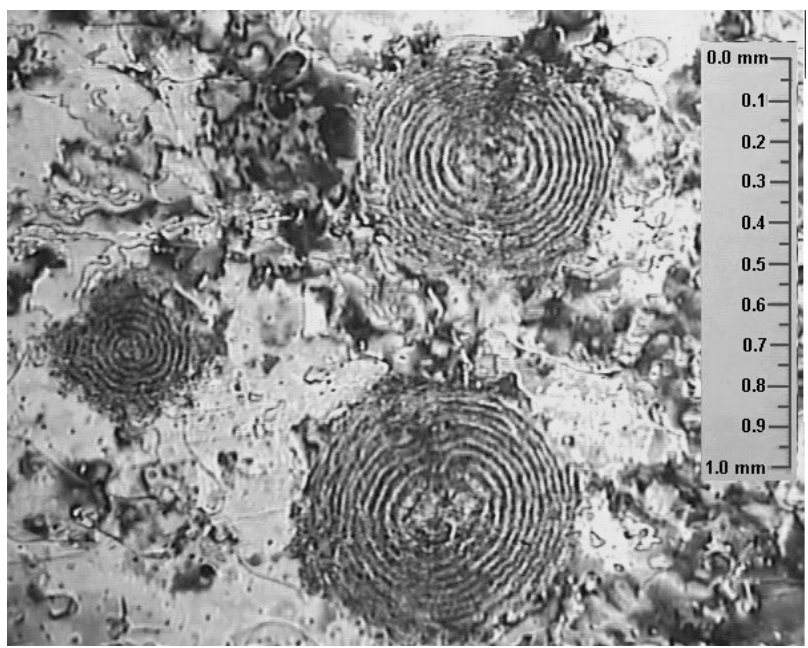

Fig.3. Spherulitic structures at room temperature (28°C) and the interference between two such structures.

\*\*\*

Fig.4. Details of the centre of spherulitic textures.

\*\*\*

Fig.5. Microphotography of a thin cell at room temperature (28°C) with crossed polarisers. Note the regions where the defects are forming a cholesteric-like texture. The lower image is the same region observed after three days. The mixture is 90% in weight of C6 and 10% of benzoic acid.

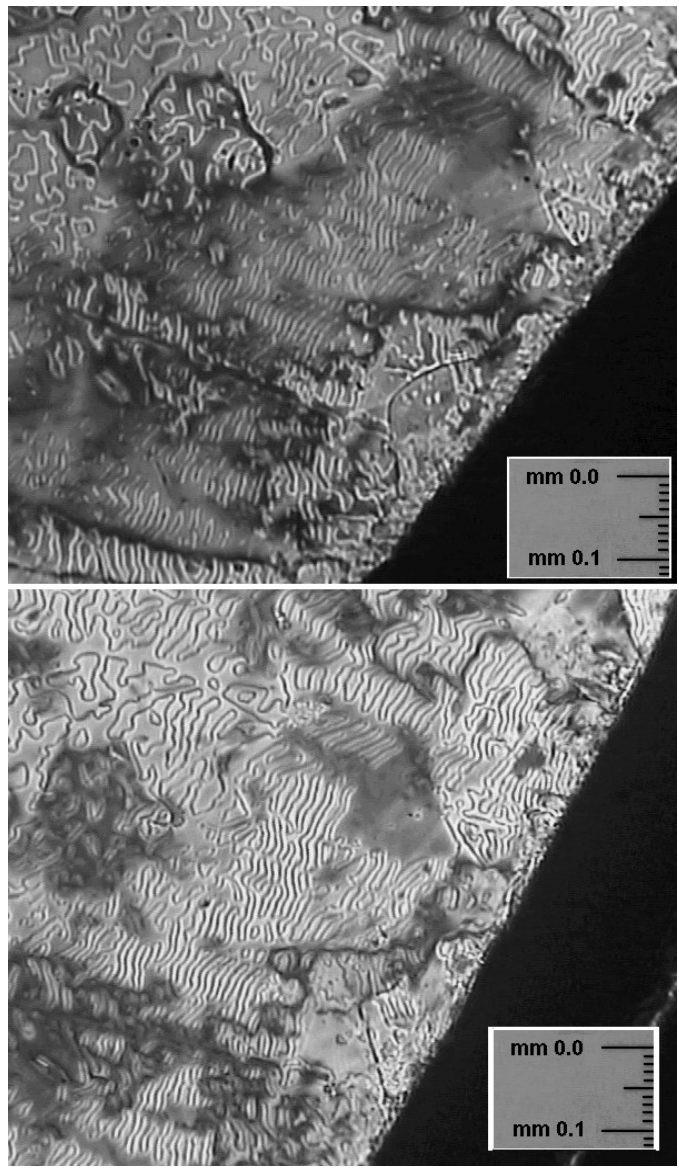

Fig.6. Microphotography of a thin cell. The observed region is the same as in Fig.5 but now the sample is rotated with respect to the crossed polarisers.

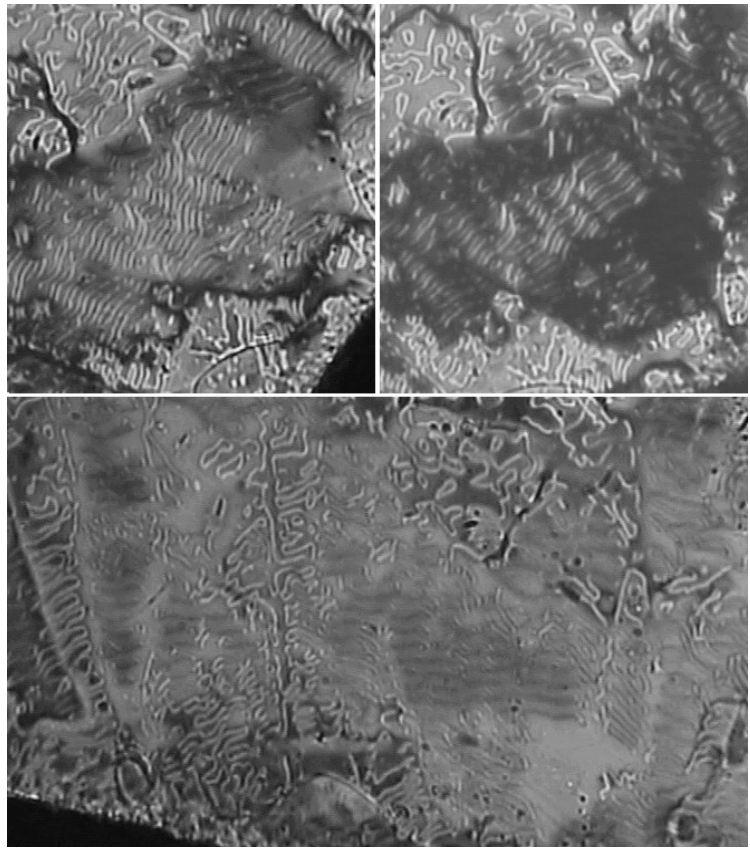

Fig.7. Behaviour of the nematic texture during the heating of the sample. Note that the upper two images display many defect lines and the texture looks like a cholesteric one. Above 71°C, defects disappear, leaving a smooth texture with few defects. On cooling, defects reappear below 70°C.

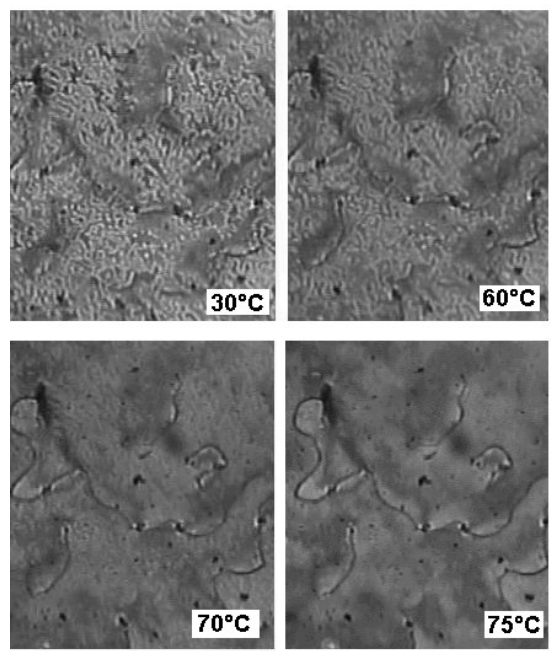

Fig.6. Microphotography of a thin cells with a binary mixture of C6 acid (95% in weight) and benzoic acid (5% in weight) at room temperature (28°C). Note the defects forming a cholesteric-like texture.

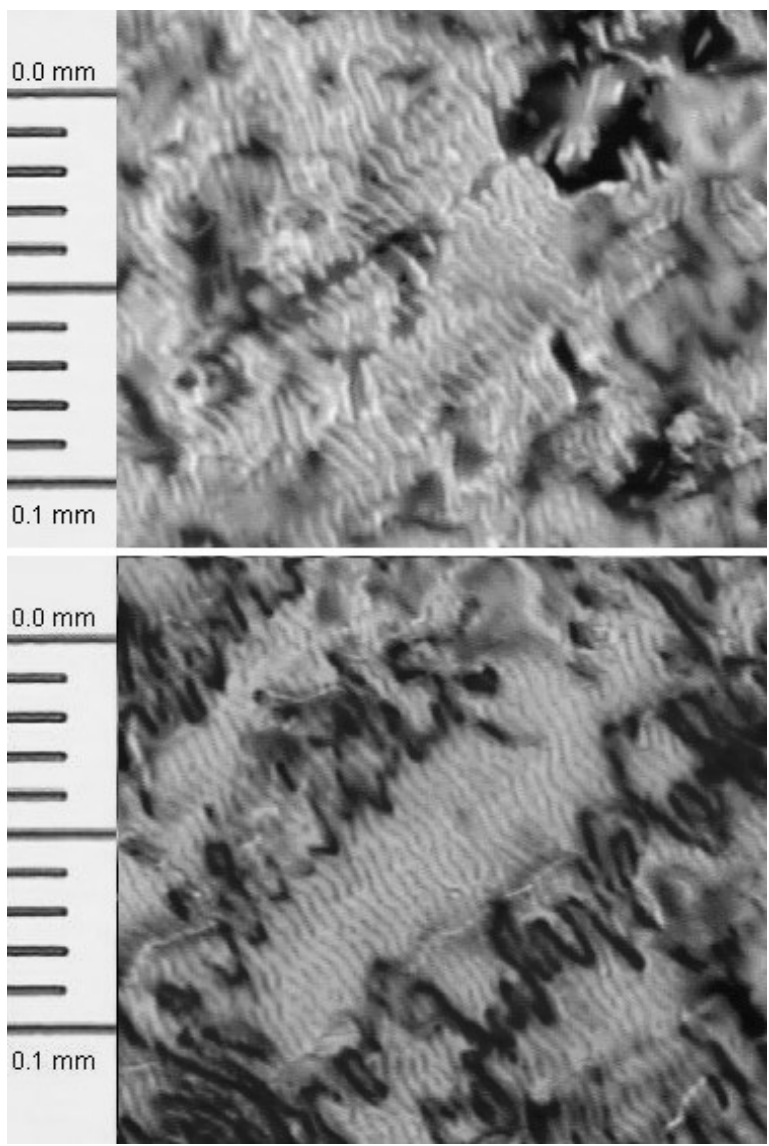